\documentstyle[12pt]{article}

\begin{document}

\begin{flushright} {OITS 621}\\
February 1997
\end{flushright}
\vspace*{1cm}

\begin{center} {\Large {\bf The Role of Gluon Depletion in
$J/\psi$ Suppression}}
\vskip .75cm
 {\bf  Rudolph C. Hwa$^\alpha$, J\'{a}n
Pi\v{s}\'{u}t$^{\alpha,\beta}$ and Neva
Pi\v{s}\'{u}tov\'a$^{\alpha,\beta}$ }
\vskip.5cm

{$^{\alpha}$Institute of Theoretical Science and Department of
Physics\\ University of Oregon, Eugene, OR 97403-5203,
USA\\
\medskip
$^{\beta}$Department of Physics, Comenius University,
SK-84215, Bratislava, Slovakia\\}
\end{center}
\vskip.5cm
\begin{abstract}
The depletion of gluons as the parton flux traverses a nucleus
in a heavy-ion collision can influence the production rate of
heavy-quark states.  Thus the suppression of $J/\psi$ can be
due to gluon depletion in the initial state in addition to nuclear
and hadronic absorption in the final state.  A formalism is
developed to describe the depletion effect.  It is shown that,
without constraints from other experimental facts beside the
$J/\psi$ suppression data in $pA$ and $AB$ collisions, it is not
possible to determine the relative importance of depletion vs
absorption. Possible relevance to the enhanced suppression
seen in the $Pb$-$Pb$ data is mentioned but not studied.

\end{abstract}

\section{Introduction} The subject of $J/\psi$ suppression in
heavy-ion collisions has been extensively investigated ever since
its first proposal as a signature of color deconfinement \cite{1}.
The recent measurement of enhanced suppression in
$Pb$-$Pb$ collisions by NA50 \cite{2} has added considerable
excitement to the possible interpretation of the data as such a
signature \cite{3,4}.  Many alternative interpretations of the data
have also been proposed \cite{5,5.1,6,6.1,6.2}.  While some of
them may have inconsistencies with all the $pA$ and nuclear
data
\cite{6.3}, as pointed out in \cite{3}, a definitive interpretation of
the $Pb$ data has not yet reached general consensus.  It is not
the purpose of this paper to add to the controversy; in fact, the
anomalous suppression in the $Pb$ data is not our main concern
here.  We want to point out that there is a loophole in
the interpretation of the $pA$ and nuclear data (prior to the
NA50 result) that is generally accepted, i.\ e.\ , the $J/\psi$
suppression is due to the absorption effects of the nuclear (and
hadronic) matter that the $c\bar{c}$ system must pass
through after it is produced.  We investigate the possibility of
another mechanism of $J/\psi$ suppression that has not been
widely considered.  It is the depletion of gluons before the
formation of the $c\bar{c}$ state that leads to $J/\psi$.  If
this new mechanism is found to be relevant to any heavy-ion
collisions, including the NA38 experiments using $O$ and $S$
beams \cite{7}, then the phenomenology of those past
experiments must be re-examined before a definite conclusion
can be reached concerning the anomalous suppression seen in
the $Pb$ data.

The essential point to be made in this paper is that what
happens to the gluons in the nuclei (apart from shadowing)
{\it before} the basic subprocess $g+g
\rightarrow c +
\bar{c}$ \ is as important as what happens to the
$c\bar{c}$ state {\it after} its formation.   Most investigations
on the subject concentrate on the latter, but the relevance of
the former can easily be seen by considering the following
extreme case.  Suppose that an ordinary nucleus $A$ collides
with an extraordinary target nucleus $B$ which is infinitely
large.  Clearly, the constituents of $A$ cannot propagate
through $B$ indefinitely without momentum degradation and
depletion.  At some penetration depth the subprocess \ $g+g
\rightarrow c +
\bar{c}$ \ just cannot take place.  Thus the production rate
of  $c\bar{c}$ (regardless of its fate afterwards) depends
on the size of $B$ and where the production points are.  If that
is accepted, then the issue becomes only a quantitative matter.
What are the sizes of $A$ and $B$ when the initial-state effects
are not negligible?

There are two aspects of the initial-state effects on the
partons:  degradation and depletion.  The degradation of parton
momenta has been suggested previously \cite{8,9}.  The
mechanism of momentum loss relies on the radiation of soft 
gluons, as the partons pass by scattering centers.  However,
such processes of multiple emission of soft gluons take time and
have been shown to be suppressed by the
Landau-Pomeranchuk-Migdal effect \cite{10}, although the energy
loss is not entirely negligible \cite{11}.  
Indeed, the  dependence of
Drell-Yan production in
$p$-$A$ collision is $\propto A^1$ \cite{15.5};
it may be taken as evidence of the ineffectiveness of
multiple small-angle scatterings of quarks.  Gluon
depletion, on the other hand, is different. 
Whereas a quark undergoing scattering must remain as a
quark, a gluon can, in addition to emitting gluons, also
create $q \bar q$ pairs as it interacts with target
partons.  When that transmutation
occurs, the gluon is lost from the beam, and
the distribution of the gluons available for $c\bar{c}$
production downstream is thereby altered.  For $J/\psi$
production the relevant momenta are $> 1.5$ GeV in the
$c\bar{c}$ rest frame.  If the subprocess of $g\rightarrow
q\bar{q}$ involves an energy change of $\Delta E > 0.8$
GeV, then the corresponding $\Delta t$ ($<0.25$ fm/c) is short
enough for the subprocess to be completed in a distance
corresponding to a mean free path $\lambda$, i.e., in $\Delta z =
\lambda/\gamma$, where $\lambda \approx 2.7$ fm, and
$\gamma$ is the Lorentz factor ($\approx 10$) for the
CERN-SPS energy.  Even if $\Delta E$ is $< 0.8$ GeV so that the
formation time for $q\bar q$ is long, those gluons that produce  
the $q\bar{q}$ upstream cannot be effective in producing $c\bar{c}$
downstream in the same nucleus. For the dominant soft processes
where the $q\bar{q}$ pairs are formed outside the nucleus, those
quarks and antiquarks cannot contribute to the production of
lepton pairs. Thus it is quite possible that the gluons can be
stripped away from the incident gluon flux, as it traverses the target
nucleus, without necessitating an enhancement of the dilepton
production rate.  Gluon depletion is therefore the loss of gluons from
the incident beam along its path for any energy change, leading to a
suppression of the $g+g\rightarrow c+\bar c$ subprocess downstream.

We know that the conversion of gluons to sea quarks
must take place efficiently in soft processes, since the gluons
that carry roughly half the incident momentum in $pp$
collisions are all turned into soft pions via the enhanced
$q\bar{q}$ sea with the same total momentum
fraction \cite{12}, while the valence quarks produce the leading
baryons, with no detectable glueballs produced.  In $pA$
collisions the wounded nucleon model that is successful in
describing soft pion multiplicities \cite{13} can be recast in the
framework of the parton model, and one obtains a picture that
is consistent with gluon depletion in that the incident gluons,
once interacted, or converted to $q\bar q$ pairs, are ineffectual
 in producing  more pions in
subsequent collisions.

The idea of gluon depletion was first applied to the problem of
$J/\psi$ suppression in $pA$ collisions by the use of an
effective gluon distribution whose deviation from the gluon
distribution in the physical nucleon increases at larger $A$
\cite{14}.  It is shown that the idea cannot be ruled out by the
existing data on $J/\psi$ and $\Upsilon$ production rates in
$pA$ collisions.  We now want to present a more detailed
analysis of the problem for
$AB$ collisions.

If gluon depletion is important in $pA$ and $AB$ collisions,
then there should be a suppression of open charm production.
There are some data on open-charm production, though not
abundant.  They are all on the single-inclusive production of
the $D$ mesons \cite{14.3}-\cite{14.6}, none on $D\bar D$ pair
production. They reveal the nuclear dependences, $A^\alpha$, that are
characterized by  values of
$\alpha$ ranging from 0.81 to 1.02.
 The uncertainty is too large to be conclusive about gluon
depletion.  For comparison the observed $A$ dependence for $J/\psi$
suppression corresponds only to $\alpha=0.92$ \cite{14.7}. 
Furthermore, since  the single-$D$ inclusive production cross-section
can include  contribution from the hadronization of the
$c\bar c$ component in the incident nucleons and from processes not
initiated by gluon fusion, those data, even if accurate, cannot
provide us with a reliable  inference on gluon depletion.  We
therefore urge dedicated experimental efforts to examine the $A$
dependence of two-particle back-to-back production of $D\bar D$.
 Information acquired in such experiments can provide
crucial constraints that can resolve some of the ambiguities uncovered
in our study in this paper.

\section{Eikonalized Gluon Depletion}

The usual expression for the production of cross section of
heavy-quark pairs $Q\bar{Q}$ is
\begin{eqnarray}
\sigma = \sum_{i,j} \int dx_1 dx_2 F_i (x_1, \mu_F) F_j (x_2,
\mu_F) \hat{\sigma}_{ij}(x_1, x_2, \mu_R)
\label{2.1}
\end{eqnarray}
where $\hat{\sigma}_{ij}(x_1, x_2, \mu_R)$ is
the cross section for the hard subprocess, $i + j \rightarrow Q +
\bar{Q}$, $i$ and $j$ being the partons involved and
$\mu_R$ being the renormalization scale.  $F_{i,j}(x_{1,2},
\mu _F)$ are the number densities of the partons $i$ and $j$ at
momentum fractions $x_1$ and $x_2$ and factorization scale
$\mu_F$.  Usually, the two scales are set equal to $\mu_R =
\mu_F = \mu = 2m_Q$.  Equation (\ref{2.1}), which is used for
hadronic collisions has generally been applied to nuclear
collisions also with the appropriate replacement of the parton
distributions by
$F_{i/A}(x_1)$ and
$F_{j/B}(x_2)$ that take into account the shadowing effects in
the nuclei $A$ and $B$.  Suppression of the detected onium
states of $Q\bar{Q}$ due to processes that take place after
the production of $Q\bar{Q}$ does not alter (\ref{2.1}),
which describes only the initial states of the hard subprocesses.

The basic point about gluon depletion is to question the validity
of (\ref{2.1}) in the gluon sector.  More specifically, the
challenge is in the factorizability of the nuclear gluon
distributions. If
$A$ and $B$ are hypothetically large, then factorization cannot
be valid on physical grounds.  We give below a formulation of
its nonfactorizability in terms of a physical cross section.

Let us denote the nuclear thickness of nucleus $A$ at impact
parameter $b_A$ by
\begin{eqnarray} T_A (b_A) = \int ^{\infty}_{-\infty} dz
\rho_A (b_A, z) \quad ,
\label{2.2}
\end{eqnarray} where $\rho_A (b_A, z)$ is the nuclear density,
normalized such that
\begin{eqnarray}
\int d^2 b_A T_A (b_A) =   A \quad .
\label{2.3}
\end{eqnarray}
For $A>1$, we define
\begin{eqnarray} T_A^- (b_A,z_A) = \left(1 -{1 \over A}\right)
\int ^{\infty}_{z_A} dz
\rho_A (b_A, z)
\quad ,
\label{2.4}
\end{eqnarray}
\begin{eqnarray} T_A^+ (b_A,z_A) = \left(1 -{1 \over A}\right)
\int ^{z_A}_{-\infty} dz
\rho_A (b_A, z)
\quad ,
\label{2.5}
\end{eqnarray}
so that $\int d^2b_A\ [T_A^+ (b_A,z_A) + T_A^- (b_A,z_A)] =
A-1$.  The variable
$z$ is positive in the direction of $A$'s momentum in the cm
system.  Thus if a hard subprocess occurs at $z_A$, then $T_A^-
(b_A,z_A)$ measures the nuclear matter in the path {\it before}
the interaction point, while $T_A^+ (b_A,z_A)$ refers to the
matter that trails {\it behind}.  Clearly, the former is relevant to
the initial-state interaction, and the latter the final-state
interaction.  Similar expressions are defined for 
$T_B^{\pm}(b_B,z_B)$, 
except that $z_B$ is positive in the opposite
direction, i.\ e.\ , in the direction of $B$'s cm momentum.
Assuming that the hard subprocess is sufficiently rare so that
in any $AB$ collision it can occur at most once, we
may identify $z_A$ with $z_B$ at the interaction point,
although the two variables are later independently integrated
over to account for all possible relative positions in the $A$ and
$B$ nuclei.

The average number of inelastic collisions that a nucleon in $A$
at $(b_A,z_A)$ suffers as it traverses $B$ in a straightline path
leading to $z_B$ is $\sigma_{\scriptsize{\mbox{in}}}T_B^-
(b_B,z_B)$, where
$\sigma_{\scriptsize{\mbox{in}}}$ is the inelastic $pN$ collision
cross section.  Strictly speaking, after the first collision of that
nucleon (call it
$p$) in $A$ with a nucleon in $B$, the former becomes a
broken nucleon (call it $p^{\prime}$) as it proceeds through the
remaining part of $B$, and the relevant cross section for the
subsequent collisions should be $\sigma^{p ^{\prime}
N}_{\scriptsize{\mbox{in}}}$ rather than
$\sigma^{pN}_{\scriptsize{\mbox{in}}}$
\cite{15}.  We do not make the distinction here now, and
denote it by the generic symbol
$\sigma_{\scriptsize{\mbox{in}}}$.  Later, it will be combined
with some other unknowns in the problem and become one
overall parameter.

The probability that $p$ makes $\nu _1$ collisions in $B$
before getting to $z_B$ is
\begin{eqnarray}
\pi _{\nu_1} (b_B,z_B) = {1 \over
\nu_1!}\left[\sigma_{\scriptsize{\mbox{in}}} T^-_B
(b_B,z_B)\right]^{\nu_1}
\mbox{exp} \left[-\sigma_{\scriptsize{\mbox{in}}} T^-_B
(b_B,z_B)\right]
\quad ,
\label{2.6}
\end{eqnarray}
where a Poisson distribution has been
assumed.  After a collision the gluon distribution is
modified by a factor $h(x_1)$, for which we assume a rather
general form
\begin{eqnarray} h(x_1) = h_0 x_1^{\alpha} (1 - x _1)^{\beta}
\quad .
\label{2.7}
\end{eqnarray}
Assuming that the same $h(x_1)$ applies at every collision, the
overall modified gluon distribution after
$\nu _1$ collisions is then
\begin{eqnarray}
G_{\nu _1}(x_1) = h^{\nu _1}
(x_1) g_A (x _1)
\quad ,
\label{2.8}
\end{eqnarray}
where $g_A (x _1)$ is the gluon distribution of a nucleon in
$A$ before any collisions but with shadowing effects taken into
account.  Similar modification takes place for the gluons in $B$
due to collisions with nucleons in $A$, and we have
\begin{eqnarray}
G_{\nu _2}(x_2) = h^{\nu _2}
(x_2) g_B (x _2)
\quad ,
\label{2.9}
\end{eqnarray}
where $\nu_2$ is the number of collisions that a nucleon in $B$
encounters in $A$ before reaching $z_A$ with probability
$\pi_{\nu_2}(b_A, z_A)$.

These modified gluon distributions are what must replace
$F_iF_j$ in (\ref{2.1}), if gluon depletion is to be taken into
account.  Thus focusing on the $gg \rightarrow Q\bar{Q}$
subprocess in (\ref{2.1}) we have for a particular interaction
point in $A$ and $B$
\begin{eqnarray}
\tilde{\sigma}_{Q \bar{Q}}\,(b_A, z_A; b_B, z_B) =
\sum^{\infty}_{\nu_1 = 0}\sum^{\infty}_{\nu_2 = 0} \pi _{\nu
_1}( b_B, z_B)\pi _{\nu
_2}(b_A, z_A)\nonumber\\
\int dx_1 dx_2 G_{\nu _1}(x_1)  G_{\nu _2}(x_2)
\hat{\sigma}_{gg \rightarrow Q\bar{Q}}(x_1, x_2)
\quad .
\label{2.10}
\end{eqnarray}
It is clear from this equation that factorization does not hold.
The gluon density at $x_1$ (in $A$) depends on the path length
in $B$ from $\infty$ to $z_B$, contained in $\pi _{\nu
_1}( b_B, z_B)$.  The total production cross section of the
$Q\bar{Q}$ state is
\begin{eqnarray}
\sigma_{Q\bar{Q}} = \int d^2 b\, d^2 s\,  d z_A\, dz_B\,
\rho_A (\vec{s}, z_A)\rho_B (\vec{s}- \vec{b}, z_B)
\tilde{\sigma}_{Q
\bar{Q}}(\vec{s}, z_A; \vec{s}-\vec{b}, z_B) \quad .
\label{2.11}
\end{eqnarray}
where $\vec{b}_A=\vec{s}$ and $\vec{b}_B=\vec{b} - \vec{s} $.
Equations (\ref{2.10}) and (\ref{2.11}) represent an
improvement of the initial-state description of $AB$ collisions
that has hitherto not been considered.

For the absorptive effects on the production of $J/\psi$ we use the
conventional description \cite{16,17}.  First, let us replace
$\hat{\sigma}_{gg
\rightarrow Q\bar{Q}} (x_1, x_2)$ in (\ref{2.10}) by the
cross section of the subprocess of $J/\psi$ production,
$\hat{\sigma}_{gg\rightarrow J/\psi} (x_1, x_2)$, before any
absorptive effect, but including intermediate states such as
$c\bar{c}g$, $\chi$, etc. \cite{17}.  Next, we take the
absorption into account by writing the survival probability in
the exponential form
\begin{eqnarray}
\mbox{exp}\left\{-\sigma _a \left[T^+_A (b_A, z_A) + T^+_B
(b_B, z_B)\right]\right\}
\label{2.12}
\end{eqnarray}
where $\sigma _a$ is the absorption cross section of $J/\psi$
interacting with the final-state medium (whether hadronic,
nuclear or quark-gluon plasma), leading to open charm.  Putting
all these factors together, we write the final result for $J/\psi$
production cross section in the following way:
\begin{eqnarray}
\sigma_{J/\psi} &=& \int d^2b\, d^2s\, dz_A\, dz_B \, \rho_A
(\vec{s}, z_A) \rho_B (\vec{s}-\vec{b}, z_B)\nonumber\\
&&\cdot \int dx_1\, dx_2\, F_A (x_1, \vec{s} - \vec{b}, z_B) F_B (x_2,
\vec{s},  z_A) \hat{\sigma}_{gg\rightarrow J/\psi} (x_1,
x_2)\nonumber\\
&&\cdot \mbox{exp} \left\{- \sigma_{\scriptsize{\mbox{in}}}
\left[T_A^-(\vec{s}, z_A) + T_B^- (\vec{s}-\vec{b}, z_B)\right] -
\sigma _a
\left[T_A^+(\vec{s}, z_A) + T_B^+ (\vec{s}-\vec{b}, z_B)
\right] \right\} .
\label{2.13}
\end{eqnarray}
where
\begin{eqnarray}
F_A (x_1, b_B, z_B) = \sum^{\infty}_{\nu_1 = 0} {1 \over
\nu_1!} \left[\sigma_{\scriptsize{\mbox{in}}}T^-_B (b_B, z_B)
\right]^{\nu_1}\, G_{\nu_1} (x_1) \quad ,
\label{2.14}
\end{eqnarray}
\begin{eqnarray}
F_B (x_2, b_A, z_A) = \sum^{\infty}_{\nu_2 = 0} {1 \over
\nu_2!} \left[\sigma_{\scriptsize{\mbox{in}}}T^-_A (b_A, z_A)
\right]^{\nu_2}\, G_{\nu_2} (x_2) \quad .
\label{2.15}
\end{eqnarray}
Equation (\ref{2.14}), for example, can be given the
interpretation of the (improperly normalized) gluon distribution
of $A$ modified by the depletion effects due to $B$.  The extra
normalization factor is the exponential term, which is now
included in the last line of (\ref{2.13}) for a reason that will
become self-evident in the next section.

\section{A Simplified Case at $x_F = 0$}

In Eq. (\ref{2.13}) we have an expression for the $J/\psi$
production cross section, obtained under a rather general
description of the gluon depletion process in high-energy
nuclear collisions.  In principle, if all the parameters controlling
different factors in the problem are known, the computation of
$\sigma_{J/\psi}$ in accordance to (\ref{2.13}) is
straightforward.  That is especially true if one determines only
the suppression effects by computing the ratio
\begin{eqnarray}
S^{AB}_{J/\psi} =
\sigma^{AB}_{J/\psi}/\sigma^{AB(0)}_{J/\psi}\quad ,
\label{3.1}
\end{eqnarray}
where $\sigma^{AB(0)}_{J/\psi}$ is the $J/\psi$ production
cross section in $AB$ collisions without absorption or depletion,
since in that case the inaccuracies in the leading-order
approximation of $\hat{\sigma}_{gg\rightarrow J/\psi} (x_1,
x_2)$ cancel in the ratio.

At this point our investigation of the gluon depletion effects is
still preliminary, since the various factors involved in the
relevant dynamics are poorly understood.  A systematic
program for its exploration should therefore begin with a
simplified calculation that can make transparent the
connections between the physics issues and their
phenomenological consequences.  More detailed calculations
can come later when proper focuses can be placed on specific
issues, after a general picture becomes clear.  Our immediate
aim is therefore to capture that general picture and see whether
gluon depletion can be relevant to the present and forthcoming
experiments in the first place.

The first step in our simplification is to consider $J/\psi$ in a 
narrow region around $x_F = 0$.  For $\sqrt{s} \simeq 20$ GeV
that means $x_1 \simeq x_2 \simeq M_{J/\psi}/\sqrt{s}
\simeq 0.15$ or slightly higher for the production of
$c\bar{c}$ state that can lead to $J/\psi$ by soft gluon
emission.  In the approximation that the integrations over
$x_1$ and $x_2$ in (\ref{2.13}) need only be extended over the
narrow range between 0.15 and, say, 0.18, beyond which
$\hat{\sigma}_{gg\rightarrow J/\psi} (x_1,
x_2)$ is negligible, we may replace the integrals by evaluating
the integrand at $x_1 = x_2 = 0.16$, and obtain
in consequence of (8) and (9)
\begin{eqnarray}
\int dx_1 \, dx_2
G_{\nu_1}(x_1) G_{\nu_2}(x_2)\hat{\sigma}_{gg\rightarrow
J/\psi} (x_1, x_2) \simeq c D^{\nu_1 + \nu_2}\quad ,
\label{3.2}
\end{eqnarray}
where $c$ and $D$ are some constants.  Actually $c$ can
depend on the nucleon numbers $A$ and $B$ on account of
nuclear shadowing, but it will be canceled in the ratio
$S^{AB}_{J/\psi}$ .  On the other hand, $D$ represents the
effect of gluon depletion and is raised to the power $\nu_1 +
\nu_2$, thereby contributing a factor of crucial importance to
us.

Using (\ref{3.2}) and ignoring $c$, we have
\begin{eqnarray}
&{}&\int dx_1 \, dx_2 \, F_A (x_1, b_B, z_B)\, F_B (x_2, b_A,
z_A)\hat{\sigma}_{gg\rightarrow J/\psi} (x_1,
x_2)\nonumber\\
 &{}&\hspace{.5in}= \sum_{\nu_1, \nu_2}{1 \over \nu_1! \nu_2! }\left[
\sigma_{\scriptsize{\mbox{in}}} T^-_B (b_B,
z_B)\right]^{\nu_1}\left[
\sigma_{\scriptsize{\mbox{in}}} T^-_A (b_A,
z_A)\right]^{\nu_2}D^{\nu_1 +
\nu_2}\nonumber\\
&{}&\hspace{.5in}=\mbox{exp}\left\{ \sigma _{\scriptsize{\mbox{in}}}
D\left[T^-_A (b_A, z_A) + T^-_B (b_B, z_B)\right]\right\} \quad .
\label{3.3}
\end{eqnarray}
Combining this simple result with the exponential factor in the
integrand in (\ref{2.13}) yields the probability factor
\begin{eqnarray}
P = \mbox{exp} \left\{- \sigma_d \left[T^-_A(\vec{s}, z_A) +
T^-_B (\vec{s}- \vec{b}, z_B)\right] - \sigma_a
\left[T^+_A(\vec{s}, z_A) + T^+_B (\vec{s}- \vec{b}, z_B)\right]
\right\}\quad ,
\label{3.4}
\end{eqnarray}
where
\begin{eqnarray}
\sigma_d = \sigma_{\scriptsize{\mbox{in}}} (1 - D) \quad .
\label{3.5}
\end{eqnarray}
We may call this the depletion cross section, inasmuch as
$\sigma_a$ is called the absorption cross section.  Note that
$\sigma_d$ is an overall parameter, summarizing a number of
imprecisely known factors.  Among them is $D$.  If there is no
gluon depletion, then in (\ref{2.7}) $h_0$ would be $1$ and
$\alpha = \beta = 0$.  In that case we would have $D = 1$ and
$\sigma_d = 0$.  In past investigations of $J/\psi$ suppression
it is universally assumed that $\sigma_d = 0$.  We now see
how nonvanishing values of $\alpha$ and $\beta$ can have
phenomenological consequences.

For phenomenology it is sufficient at first to start with
(\ref{3.4}), which involves just two parameters, $\sigma_d$
and $\sigma_a$.  The thickness functions $T^{\pm}_{A,B}$
accompanying $\sigma_d$ and $\sigma_a$ are just what they
should be.  $T^-_A$ is the nuclear thickness of $A$ {\it before} the
interaction point, and $T^-_B$ is that for $B$, while
$T^{+}_{A,B}$ are the respective thicknesses {\it after} the
interaction point.  The symmetry of the suppression mechanisms
is now complete:  depletion during the pre-interaction phase
and absorption during the post-interaction phase.  Without
further study it is not obvious which is more important.

Integration over the geometrical variables can be significantly
simplified without much sacrifice in accuracy, if we
approximate the nuclear density by a constant value $\rho_0$,
i.e., $\rho_{A,B}(r) = \rho_0 \Theta (R_{A,B} - r)$ where $R_A$
and $R_B$ are the radii of $A$ and $B$, respectively.  In that
approximation the nuclear thicknesses are
\begin{eqnarray}
T^\pm_{A} = \rho^0_A (L_{A} \pm z_{A})\quad ,\quad \qquad \rho^0_A =
\left(1-1/A \right)\rho_0 \quad ,
\label{3.6}
\end{eqnarray}
and similarly for $T^{\pm}_B$, where
\begin{eqnarray}
L_A = (R_A^2 - s^2)^{1/2}\quad, \quad \quad L_B =  (R^2_B
-\left|\vec{s}- \vec{b} \right|^2)^{1/2}\quad .
\label{3.7}
\end{eqnarray}
Combining (\ref{2.13}),  (\ref{3.1}) and (\ref{3.4}) we have for
the suppression factor
\begin{eqnarray}
S^{AB}_{J/\psi} = N_{AB}^{-1} \int d^2b \, d^2s \, U(\vec{b},
\vec{s})
\label{3.8}
\end{eqnarray}
where
\begin{eqnarray}
U(\vec{b},\vec{s}) 
& = & \int^{L_A}_{-L_A}dz_A
\int^{L_B}_{-L_B}dz_B \,\mbox{exp} \left\{  - \sigma_d
\left[\rho^0_A \left(L_A - z_A\right) + \rho^0_B\left(L_B -
z_B\right)\right]\right.\nonumber\\
& & \left. \hspace{4cm} \mbox{}-\sigma_a
\left[\rho^0_A \left(L_A + z_A\right) + \rho^0_B \left(L_B +
z_B \right)\right]\right\}\nonumber\\ 
& = &\left( e^{-2\sigma_a\rho^0_A
L_A} - e^{-2\sigma_d \rho^0_A L_A}\right)\left(
e^{-2\sigma_a\rho^0_B L_B} - e^{-2\sigma_d\rho^0_B
L_B}\right)/\left[\rho^0_A\rho^0_B\left( \sigma_d -
\sigma_a\right)^2 \right]\   ,
\label{3.9}
\end{eqnarray}
\begin{eqnarray}
N_{AB}&=&4 \int d^2b \, d^2s\, L_a(s) L_B
\left(\left|\vec{s}- \vec{b} \right|\right) \quad .
\label{3.10}
\end{eqnarray}
The symmetry of the problem under the interchange of
$\sigma_a$ and $\sigma_d$ is now explicit.  Furthermore, if
$\sigma_d = 0$, (\ref{3.8}) and (\ref{3.9}) agree with the
corresponding formula derived by Gerschel and H\"{u}fner
\cite{16}.

Since we know very little about the dynamics of gluon
depletion, we have no reliable information on the magnitude of
$\sigma_d$.  However, we do know that the Gerschel and
H\"{u}fner formula can fit the heavy-ion data on $J/\psi$
suppression, excluding the $Pb$-$Pb$ collision result \cite{2}, by
use of $\sigma_a = $
6-7 mb 
(and, of course $\sigma_d = 0$)
\cite{3}.  We therefore can expect that when the effects of
gluon depletion are considered, the combined suppression
mechanisms would have roughly the same overall cross section.
Nevertheless, we use the combined cross section defined by
\begin{eqnarray}
\sigma_{c} = \sigma_a + \sigma_d
\label{3.11}
\end{eqnarray}
as a free parameter to fit the pre-$Pb$ data.
The ratio $\eta \equiv \sigma_d/\sigma_a$ can still vary
between 0 and 1.  The range $\eta > 1$ leads to no new result
because of the $\sigma_a \leftrightarrow \sigma_d$ symmetry
of $U(\vec{b}, \vec{s})$.  Thus the measurement of
$S^{AB}_{J/\psi}$ cannot resolve this ambiguity at this level of
consideration.  For definiteness we shall examine the range $0
< \eta < 1$.   In that range it is not obvious by inspecting
(\ref{3.8})-(\ref{3.10}) how $S^{AB}_{J/\psi}$ depends on $A$
and $B$.  A numerical computation is therefore necessary.

\section{Some Numerical Results}

A crude but quick estimate of (\ref{3.8}) without doing the
integrations is to replace $L_A$ and $L_B$ in the integrands by
their averages, ${3 \over 4}R_A$ and ${3 \over 4}R_B$,
respectively \cite{16}, and $\rho^0_{A,B}$ by $\rho_0$ for $A$
and $B > 1$.  Denoting the resultant approximation of
$S^{AB}_{J/\psi}$  by
$\bar{S}^{AB}_{J/\psi}$, one obtains the analytic form
\begin{eqnarray}
\bar{S}^{AB}_{J/\psi} = { 4
\over 9 R_A R_B (\lambda^{-1}_d - \lambda^{-1}_a)^2} \left(
e^{-{3R_A
\over 2
\lambda_a}}-e^{-{3R_A
\over 2 \lambda_d}}\right)\cdot \left( e^{-{3R_B \over 2
\lambda_a}}-e^{-{3R_B
\over 2 \lambda_d}}\right)\quad ,
\label{4.1}
\end{eqnarray}
where $\lambda_a$ and $\lambda_d$ are the mean free paths
\begin{eqnarray}
\lambda_a = (\sigma_a\rho_0)^{-1}, \qquad \lambda_d =
(\sigma_d\rho_0)^{-1} \quad .
\label{4.2}
\end{eqnarray}
For $pA$ collisions (\ref{4.1}) becomes
\begin{eqnarray}
\bar{S}^{pA}_{J/\psi} = { 2
\over 3 R_A  (\lambda^{-1}_d - \lambda^{-1}_a)} \left(
e^{-{3R_A
\over 2
\lambda_a}}-e^{-{3R_A
\over 2 \lambda_d}}\right) \quad .
\label{28`}
\end{eqnarray}
Using $R_A = 1.2A^{1/3}$ fm and $\rho^{-1}_0 = {4 \over 3}
\pi (1.2)^3$ fm$^3$, (\ref{4.1}) and (\ref{28`}) can be 
calculated as a
function of $A$ and $B$ for various values of $\eta =
\sigma_d/\sigma_a$ subject to the constraint (\ref{3.11}).
The numerical result for that will be given and discussed
below.

A direct computation of $S^{AB}_{J/\psi}$ by carrying out the
integrations in (\ref{3.8}) is, of course, straightforward.  Since
the measurable total transverse energy $E_T$ depends on $b$,
the integration in $\vec{b}$  can be suspended by plotting the
suppression factor against the mean longitudinal length $L(b)$
at fixed $b$, which in turn can be related to the $E_T$.  $L(b)$
is defined by
\begin{eqnarray}
L(b) &=& \left< L_A + z_A + L_B +z_B\right>_{\vec{s}, z_A,
z_B}\nonumber\\
&=&{\int^{R_A}_0 sds \int^{2\pi}_0 d\theta L_A(s)
L_B(\vec{b}-\vec{s})\left[ L_A(s) +
L_B(\vec{b}-\vec{s})\right] \over \int^{R_A}_0 sds \int^{2\pi}_0 d\theta L_A(s)
L_B(\vec{b}-\vec{s})} \quad ,
\label{4.3}
\end{eqnarray}
where
\begin{eqnarray}
L_B(\vec{b}-\vec{s})=\left(R^2_B - b^2- s^2 + 2bs \cos \theta
\right)^{1/2} \Theta \left(R_B - \left| \vec{b}-\vec{s} \right|
\right) \quad ,
\label{4.4}
\end{eqnarray}
$\Theta$ being the step function.  Parenthetically, we remark
that the value mentioned earlier for the average $L_A$ is
\begin{eqnarray}
\left<L_A\right>_{s,z_A} = \left({4 \over 3} \pi R^3_A
\right)^{-1}\int d^2s \int^{L_A}_{-L_A} dz_A \left[L_A (s) +
z_A\right] = {3 \over 4} R_A \quad .
\label{4.5}
\end{eqnarray}
Thus the approximation made in (\ref{4.1}) is effectively
 \begin{eqnarray}
\left<L(b)\right>_b \simeq \left<L_A\right>_{b_A,z_A} +
\left<L_B\right>_{b_B,z_B} \quad .
\label{4.6}
\end{eqnarray}

The calculated results on $\bar S^{AB}_{J/\psi}$ obtained from the
use of (27) or (29), as the case may be, and on the more precise
$S^{AB}_{J/\psi}$ from (23)-(25) turn out to be very nearly the same
with differences being at the level of $<1\%$.  We therefore present a
single figure to represent both.  It is shown in Fig.\ 1. The
triangles are the calculated points for $\sigma_c=6$ and 7 mb, and
$\eta=\sigma_d/\sigma_a=0.05.$  As expected, they compare very well
with the data \cite{2} as shown in open squares, except for the
$Pb$-$Pb$ point.  The solid straight lines are best fits of the
triangles, and provide adequate fits of the data up to $SU$.  That
is the known result from earlier work before the $Pb$ data, and is
based on no gluon depletion.  Now we fix $\sigma_c$ at 7 mb
and increase $\eta$ to 1.  The result is shown by the circles, which
are fitted by the dotted line according to a quadratic formula of
the form $a+bx+cx^2$.  Although the result does not differ too much
from the $\eta=0.05$ case, there is nevertheless a perceptible bend
downward at higher values of $AB$. We find that to be a very
encouraging sign for the possible interpretation of the $Pb$ data as
being a manifestation of the gluon depletion effect, provided that
an enhancement at high values of $AB$ can be incorporated in an
improved description of the effect.

In Fig.\ 2 we show the results for $S^{AB}_{J/\psi}$ calculated at
specific values of $b$, but are plotted against $L(b)$ using (30). 
The same values of $\sigma_c$ and $\eta$ as in Fig.\ 1 are used. 
The general agreement with the data is again very similar to that in
Fig.\ 1, as it should.  The fact that the $\eta=1$ points  get
within the error bars of two of the three $Pb$-$Pb$ data points
provides further motivation to take the possibility of gluon
depletion seriously.

The most striking feature in those Figs.\ 1 and 2 is that the
dependences on $\eta$ are very small.  Without
further physics inputs it is not possible to extract from the data
the realistic value of $\eta$.  A conservative conclusion is
therefore that the possible contribution of gluon depletion to
$J/\psi$ suppression cannot be ruled out.  This statment is already
of considerable importance in our view because firstly it means that a
loophole has been found in the conventional approach to
$J/\psi$ suppression and secondly with a crack opened in this
new way of considering the suppression mechanisms it is
possible to imagine enhanced suppression at large $A$ and $B$,
as will be discussed in the following section.  The downturn of
the $Pb$ data requires an increase of either $\sigma_a$ or
$\sigma_d$.  Any speculation on the
increase of $\sigma_d$ is not the purpose of this paper.
However, until that possibility is firmly ruled out, the increase
of
$\sigma_a$ as the explanation of the $Pb$ data should only be
held as tentative, albeit a very attractive one.

\section{Comments}

We have raised the issue of gluon depletion in heavy-ion
collisions, developed a formalism to describe its effects on
$J/\psi$ production, performed numerical computation to
examine its consequences, and shown that the present data
cannot exclude its possible contribution to $J/\psi$
suppression.  The combined cross section is found to be 
$\sigma_{c} \simeq 7$ mb, 
but the ratio $\eta =
\sigma_d/\sigma_a$ is undetermined because of the
insensitivity of the suppression factor to $\eta$.

An independent experimental constraint is necessary to
determine $\eta$.  We suggest the $A$ dependence of back-to-back
correlated production of $D\bar D$ in $pA$ collisions.  For $D\bar{D}$
production near threshold the formalism that we have
developed is applicable, if $\sigma_a$ is set to zero.  Thus any
observation of $\sigma_{D\bar{D}} \propto A^{\alpha}$
with $\alpha < 1$ would be a signature of gluon depletion.

It is also possible to get extra information from $J/\psi$
suppression if we examine the $x_F$ dependence.  The
restriciton to $x_F = 0$ in Sec.\ 3 simplifies the problem so that
the suppression factor $S^{AB}_{J/\psi}$ does not depend on the
detailed gluon distributions.  However, for $x_F \neq 0$, the
details of all the factors in (\ref{2.10}) will become relevant,
and the effect of gluon depletion cannot be described by one
collective parameter $\sigma_d$.  Thus the subject
has the potential of developing into a fertile field of
phenomenology.

A more pressing question is perhaps inescapable:  does gluon
depletion have any relevance to the more urgent issue of
enhanced suppression observed in NA50 \cite{2}?  We have
avoided addressing that issue in order to be clear about what
constraints the pre-NA50 data can place on gluon
depletion, the pertinence of which in heavy-ion collisions
should be investigated independent of the NA50 data.  We
now ask whether there is any chance that the downturn of
the suppression factor in the $Pb$-$Pb$ data can be due to
some aspects of the gluon depletion process.  The formalism
described in Secs. 2 and 3 does not lead to any
prominent nonlinear behaviors in Figs.\ 1 and 2.  
However, it should
be noted that the modified gluon distribution in (\ref{2.8}) is
obtained under the assumption that the same depletion factor 
$h(x_1)$
applies at each of the $\nu_1$ collisions.  That is a reasonable
first-try to estimate the effect of multiple collisions, but it
does not follow from any careful dynamical consideration.  In
the absence of a workable nonperturbative QCD calculation
one can envisage a study in which gluon depletion is viewed as
a gain-loss evolution process where gluons in a cell of
momentum fraction $x_1 \sim 0.15$ are lost from the cell due
to the $g \rightarrow q\bar{q}$ subprocess, but the gain
comes from higher-$x_1$ cells due to $g \rightarrow gg$, for
example.  Since the initial gluon distribution $g_A(x_1)$
behaves roughly as $(1-x_1)^5$, there are less gains than losses
as $\nu_1$ increases.  Thus it is conceivable that the
modification factor in front of $g_A(x_1)$ in (\ref{2.8}) may
increase as a nonlinear power of  $\nu_1$, resulting in an
enhanced effect at higher $\nu_1$.  The overall suppression
when combined with absorption may show a break from
linearity if below the crossover point the absorption is
dominant, while above it the depletion is more important.
These are speculations that need concrete calculations to gain
substance.  Nevertheless, unless such possibilities are excluded,
there is no water-tight argument in favor of any other
interpretation of the enhanced suppression.

Despite the alluring challenges posed by the NA50 data, we feel
that at this point it is more important to pin down the extent of
gluon depletion in a systematic way, with emphases on $pA$
collisions, correlated $D\bar{D}$ production, and the longitudinal
momentum dependence of $J/\psi$ production.

\section*{Acknowledgments}We have benefitted from
discussions with R.\ Lietava, H.\ Satz and X.-N.\ Wang.  This work
was supported, in part, by the U.\ S.\ - Slovak Science and
Technology Program, the National Science Foundation under
Grant No. INT-9319091 and by the U.\ S.\ Department of
Energy under Grant No. DE-FG03-96ER40972.

\vspace{2cm}
\begin{center}
\section*{Figure Captions}
\end{center}
\begin{description}

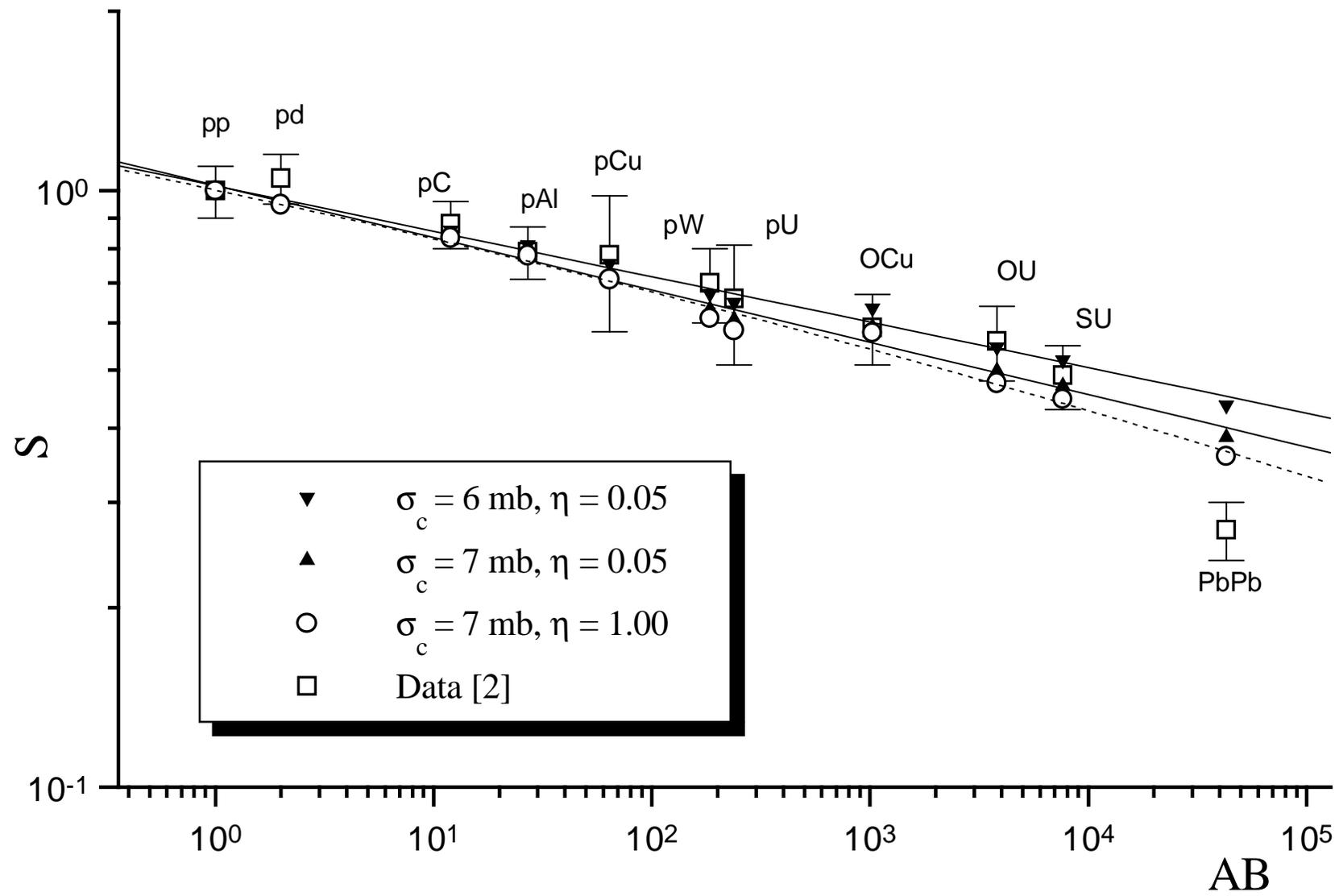
\item[Fig.\ 1]\quad The suppression factor $S^{AB}_{J/\psi}$,
abbreviated as $S$, is plotted against $AB$ for various
combinations of the values of the combined cross-section $\sigma_c$
and the ratio
$\eta=\sigma_d/\sigma_a$. The data are from \cite{2}.

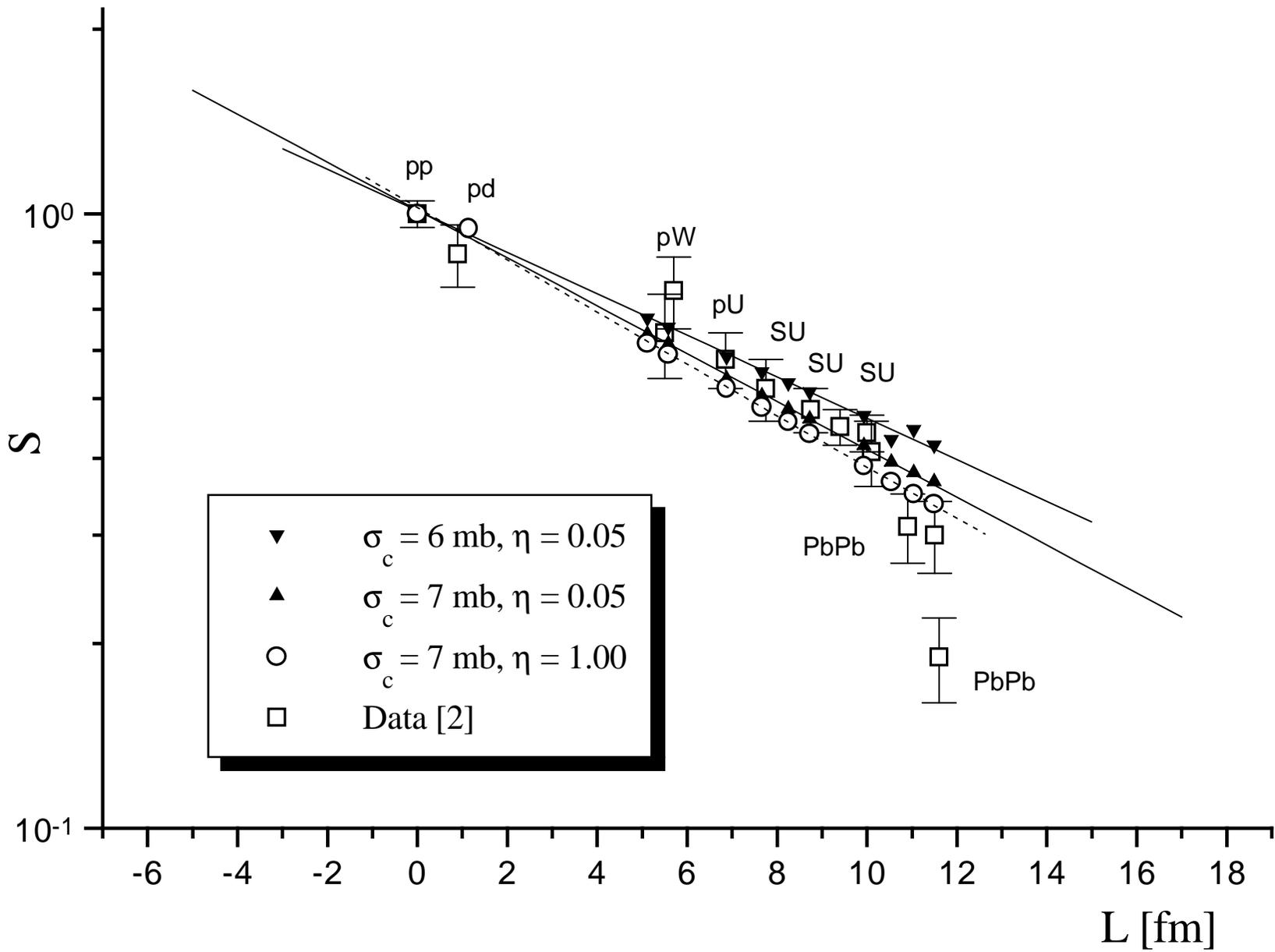
\item[Fig.\ 2]\quad Same as in Fig.\ 1, but plotted against $L(b)$.

\end{description}


\begin{thebibliography}{99}

\bibitem{1}T.\ Matsui and H.\ Satz,  Nucl.\ Phys.\ B{\bf
326}, 613 (1989).

\bibitem{2}M.\ Gonin, (NA50), {\it Quark Matter '96}, Heidelberg.


\bibitem{3}D.\ Kharzeev, C.\ Louren\c{c}o, M.\ Nardi, and H.\
Satz, CERN-TH/96-328, BI-TP 96/53, hep-ph/9612217.

\bibitem{4}J.-P.\ Blaizot and J.-Y.\ Ollitrault,  Phys.\
Rev.\ Lett. {\bf 77}, 1703 (1996).

\bibitem{5}S.\ Gavin and R.\ Vogt, hep-ph/9606460;
nucl-th/9609064.

\bibitem{5.1}A.\ Capella, A.\ Kaidalov, A.\ Kouidu Akil, and C.\
Gerschel, hep-ph/9607265.

\bibitem{6}J.\ Ft\'{a}\u{c}nik, J.\ Pi\v{s}\'{u}t,
and N.\ Pi\v{s}\'{u}tov\'{a}, Comenius preprint (Sept.
1996).

\bibitem{6.1}C.\ Y.\ Wong, hep-ph/9607285.

\bibitem{6.2}W.\ Cassing and C.\ M.\ Ko, nucl-th/9609025.

\bibitem{6.3}C.\ Louren\c{c}o, in  {\it Quark Matter 1996},
Heidelberg (to be published).


\bibitem{7}C. Baglin {\it et al.} (NA38),  Phys.\ Lett. B {\bf
220}, 471 (1989); B {\bf 255}, 459 (1991); B {\bf 270}, 105 (1991).

\bibitem{8} S.\ Gavin and J.\ Milana,  Phys.\ Rev.\ Lett.
{\bf 68}, 1834 (1992)

\bibitem{9}E.\ Quack and T.\ Kodama,  Phys. Lett. B {\bf
302}, 495 (1993).

\bibitem{10}L.\ D.\ Landau and I.\ Ya.\ Pomeranchuk,  Dokl.
Akad. Nauk SSSR {\bf 92}, 535, 735 (1953); A.\ B.\ Migdal, 
Phys.\ Rev.\ {\bf 103}, 1811 (1956); E.\ L.\ Feinberg and I.\ Ya.\
Pomeranchuk,  Supp.\ Nuovo Cimento {\bf 3}, 652 (1956).

\bibitem{11} M.\ Gyulassy and X.-N.\ Wang,  Nucl.\ Phys.\ B
{\bf 420}, 583 (1994); X.-N.\ Wang, M.\ Gyulassy and M.\ Pl\"umer,
 Phys.\ Rev.\ D {\bf 51}, 3436 (1995).

\bibitem{15.5} D.\ M.\ Alde  et al., Phys.\ Rev.\ Lett.\ {\bf
64}, 2479 (1990).

\bibitem{12}   R.\ C.\ Hwa,  Phys.\ Rev. D
{\bf 22}, 1593 (1980).

\bibitem{13} A.\ Bia\l as, M.\ B\l eszy\'{n}ski, and W.\ Czy\.{z},
 Nucl.\ Phys. B {\bf 111}, 461 (1976); A.\ Bia\l as, W.\
Czy\.{z} and W.\ Furm\'{a}nski,   Acta Phys.\ Polon.\ 
B{\bf 8}, 585 (1977).

\bibitem{14}  R.\ C.\ Hwa and L.\ Le\'{s}niak,   Phys. Lett.
B {\bf 295}, 11 (1992).

\bibitem{14.3} H. Cobbaert {\it et al.} (WA78 Collaboration), Phys.
Lett. B {\bf 191}, 456 (1987).

\bibitem{14.4} M.\ Adamovich {\it et al.} (WA82 Collaboration),
Phys.\ Lett.\ B {\bf 284}, 453 (1992).

\bibitem{14.5} G.\ A. Alves {\it et al.} (E769 Collaboration), Phys.\
Rev.\ Lett.\ {\bf 70}, 722 (1993).

\bibitem{14.6} M.\ J.\ Leitch {\it et al.},  Phys.\ Rev.\ Lett.
{\bf 72}, 2542 (1994).

\bibitem{14.7} D.\ M.\ Alde {\it et al.}, Phys.\ Rev.\ Lett.\ {\bf
66}, 133 and 2285 (1991).

\bibitem{15}  R.\ C.\ Hwa and W.\ N.\ Wang,  Phys.\ Rev.
D {\bf 39}, 2561 (1989).

\bibitem{16}C.\ Gerschel and J.\ H\"{u}fner,   Phys. Lett.
B {\bf 207}, 253 (1988);   Z.\ Phys.\ C
 {\bf 56}, 171 (1992).

\bibitem{17}D.\ Kharzeev and H.\ Satz, in {\em Quark-Gluon
Plasma 2}, edited by R.\ C.\ Hwa (World Scientific, Singapore,
1996).

\end{thebibliography}
\end{document}